 \documentclass[manuscript]{emulateapj}

\slugcomment{to appear in the Astrophysical Journal}

\shorttitle{Light Curve Analysis of V838 Her}
\shortauthors{Kato et al.}

\begin{document}


\title{A Universal Decline Law of Classical Novae. IV.  V838 Her (1991):
A Very Massive White Dwarf}


\author{Mariko Kato}
\affil{Department of Astronomy, Keio University, Hiyoshi, Yokohama
  223-8521, Japan  }
\email{mariko@educ.cc.keio.ac.jp}

\author{Izumi Hachisu}
\affil{Department of Earth Science and Astronomy, College of Arts and
Sciences, University of Tokyo, Komaba, Meguro-ku, Tokyo 153-8902, Japan}
\email{hachisu@ea.c.u-tokyo.ac.jp}

\and

\author{Angelo Cassatella}
\affil{INAF, Istituto di Fisica dello Spazio Interplanetario,
 Via del Fosso del Cavaliere 100, 00133 Roma, Italy}
\affil{Departamento de Astrof\'isica, Facultad de F\'isica, Universidad
  Complutense de Madrid, 28040 Madrid, Spain} 
\affil{Dipartimento di Fisica E. Amaldi, Universit\`a degli Studi Roma Tre, 
Via della Vasca Navale 84, 00146 Roma, Italy}
\email{cassatella@fis.uniroma3.it}




\begin{abstract}
We present a unified model of optical and ultraviolet (UV) light curves for
one of the fastest classical novae, V838 Herculis (Nova Herculis 1991), and
estimate its white dwarf (WD) mass. Based on an optically thick wind theory 
of nova outbursts, we model the optical light curves with free-free emission
and the UV 1455 \AA~ light curves with blackbody emission.  Our 
models of $1.35 \pm 0.02~M_\odot$ WD reproduce simultaneously the optical
and UV 1455 \AA~ observations.  The mass lost by the wind
is $\Delta ~M_{\rm wind} \sim 2 \times 10^{-6} ~M_\odot$.  We provide new
determinations of the reddening, $E(B-V) = 0.53 \pm 0.05$, and of the distance, 
$2.7 \pm 0.5$~kpc.
\end{abstract}





\section{Introduction}
\label{introduction}

V838 Her was discovered independently by \citet{sug91} on 1991 March 24.78 UT
at 5.4 mag and by \citet{alc91} at $\sim 5$ mag on 1991 March 25.67 UT.
Alcock's visual discovery around the peak was made in strong twilight, so the
peak magnitude was not accurate.  The outburst time can be estimated
from the upper limit (9 mag) prediscovery observation by \citet{uet91}
on JD 2,448,339.9 and Sugano's discovery on JD 2,448,340.281.  In the present
work we adopt the outburst time JD~2,448,340.0 as day zero and the
maximum magnitude $m_{\rm V, max}=5.4$ mag.

Soon after the optical maximum, the nova entered a very rapid
  decline phase, followed by a slight oscillatory behavior enduring several
days and, later on, by a smooth decline.  Infrared (IR) fluxes also showed a
rapid decline, reaching a local minimum on day 6, followed by a rapid
brightening.  The spectral energy distribution was consistent with free-free
emission from day 1.3 to 6.5, and thereafter with blackbody emission, ascribed
to the formation of a hot dust shell \citep{cha92,har94,kid93,woo92}.  Indication of
IR emission from silicate grains was reported only at a later phase
\citep{lyn92,smi95}.  \citet{smi95} suggested that silicate emission was due
to a light echo by cold silicate grains deposited in a previous nova eruption.

Red and blue preoutburst magnitudes were estimated from the Palomar Sky Survey
plates as 19 and 17.5, respectively, by \citet{wes91}, and 20.6 and 18.25
by \citet{hum91}, but were contaminated by a closeby ($\sim 1''$) star
of similar magnitude. V838 Her returned to the preoutburst magnitude
$V=19.0$ on day 403, and $V=19.2$ on day 572 \citep{szk94i}.

In this series of papers, we have applied our universal decline law 
to many novae, including moderately fast novae (e.g., 
V1668 Cyg in \citet{hac06a} [hereafter referred as Paper I]) 
and a slow nova (GQ Mus in \citet{hac08}[ Paper III]), and have 
succeeded in reproducing simultaneously 
optical, infrared, ultraviolet (UV) 1455 \AA, and supersoft X-ray light curves. 
V838 Her is, however, one of the fastest classical novae, so it is 
interesting to see if the universal decline law also applied to such 
an extreme object.  
The main observational features of V838 Her are summarized in Tables
\ref{table_observation} and \ref{table_v838her_chemical_abundance}. 
Section \ref{sec_UV} reviews
our observational results based on the {\it IUE} spectra.  Our model light curves 
are briefly introduced in Section \ref{sec_model}. The results of light 
curve fitting are
given in Section \ref{sec_lightcurvefitting}.  Discussions and conclusions follow
in Sections \ref{sec_discussion} and \ref{sec_conclusion}.  In Appendix 
we give several examples of nova light curves, which show high degree 
of homology, as predicted by our models.


\begin{deluxetable*}{llll}
\tabletypesize{\scriptsize}
\tablecaption{Observational Properties of V838 Herculis
\label{table_observation}}
\tablewidth{0pt}
\tablehead{
\colhead{subject} &
\colhead{} &
\colhead{data} &
\colhead{reference}
}
\startdata
discovery & ... & JD 2,448,340.28 &  \citet{sug91} \\
nova speed class & ... & very fast &          \\
IR minimum & ... & JD 2,448,345.94 (day 6)  &       \\
$t_3$ & ... & 5.3-6 days & \citet{har94} \\
$m_{v, {\rm max}}$  &... & 5.4 mag & \citet{sug91}\\
$M_{V, {\rm max}}$ from $t_3$ &... & $-9.94 \sim -9.80$ mag & equation (\ref{schmidt-kaler-law})\\
distance &... & 2.8-10 kpc &  \citet{van96,har94} \\
distance &... & $2.7 \pm 0.5$ kpc &  this work \\
FWHM of UV 1455 \AA &... & 4.7 days& this work \\ 
$E(B-V)$ & ... &0.3-0.6 &  \citet{van96,har94}\\
$E(B-V)$ & ... &$0.53 \pm 0.05$  &  this work\\
dust & ... & very thin &     see \S \ref{sec:dust} \\
orbital period & ... & 7.14 hr & \citet{ing92,lei92} \\
H-burning phase & ... & $< 1$  yr & \citet{szk94h}
\enddata
\end{deluxetable*}


%
\begin{deluxetable*}{llllllll}
\tabletypesize{\scriptsize}
\tablecaption{Chemical Abundance by Weight
\label{table_v838her_chemical_abundance}}
\tablewidth{0pt}
\tablehead{
\colhead{object} &
\colhead{$X$} &
\colhead{$Y$} &
\colhead{$CNO$} &
\colhead{$Ne$} &
\colhead{$Na-Fe$} &
\colhead{reference}
}
\startdata
Sun &   0.7068 &0.274    &0.014  &0.0018 &0.0034 & \citet{gre89} \\
V838 Her & 0.78 &0.10    &0.041  &0.081  &0.003& \citet{van96} \\
V838 Her & 0.59 &0.31    &0.030  &0.067  &0.003 & \citet{van97} \\
V838 Her & 0.562 & 0.314  &0.038 & 0.070 & 0.015 & \citet{sch07}
\enddata
\end{deluxetable*}

\section{UV observations } 
\label{sec_UV}

V838 Her has been monitored by {\it IUE} from day 2 to day 831, 
mainly at low resolution. A gallery of UV spectra can be found in
\citet{cas04a} and \citet{van96}.

In the following we revisit the problem of color excess $E(B-V)$ of
V838 Her, and describe the long term evolution of UV continuum and
of emission lines.  The UV spectra were retrieved from the {\it
  IUE} archive through the INES ({\it IUE} Newly Extracted Spectra)
system\footnote{http://sdc.laeff.inta.es/ines/}, 
which also provides full details of
the observations.  The use of the {\it IUE} INES data is particularly important
for the determination of reddening correction because of the
implementation of upgraded spectral extraction and flux calibration procedures
compared to the previously published UV spectra.

\subsection{Reddening Correction}

As summarized by \citet{van96}, the color excess of V838 Her has been
determined by several authors using different methods, based on the
Balmer decrement \citep{ing92,van96}, the equivalent width of the Na I
interstellar lines \citep{lyn92}, the ratio of the UV flux above and below 2000
\AA~ \citep{sta92}, and the assumed intrinsic color at maximum light
\citep{woo92}.  The mean value of these determinations, assuming to have the
same weight, is $E(B-V)$ = 0.50 $\pm$ 0.12. Such a large error would cause
large uncertainties on the distance and hence on the intrinsic parameters of
the nova, as derived through a comparison with our models. This has
driven us to revisit the problem of the reddening using an independent
method that we consider more reliable than the previous ones, because of 
relying directly on its best detectable effect, i.e. the strength and shape
of the 2175 \AA~ broad dust absorption band.

\begin{figure}
\epsscale{1.1}
\plotone{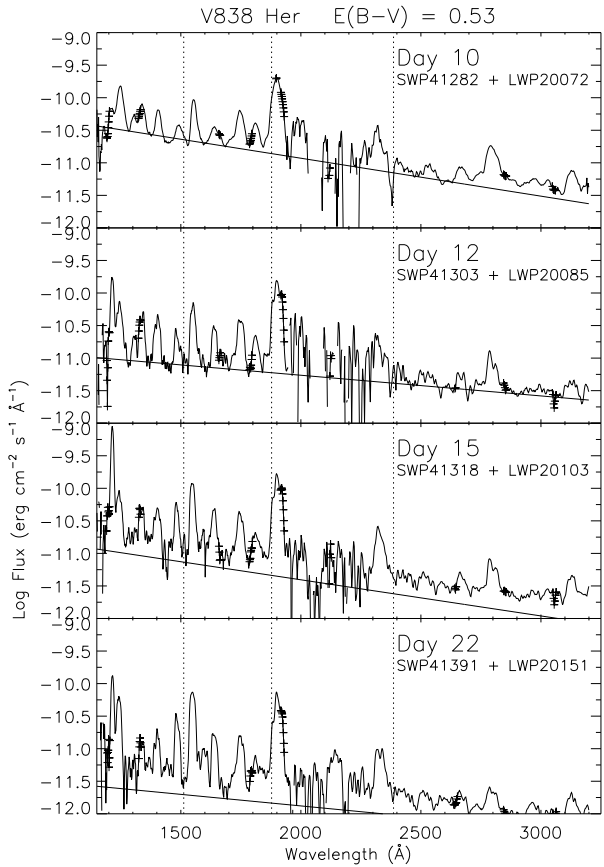}
\caption{{\it IUE} spectra of V838 Her obtained at different dates
(days after the outburst).
The spectra have been corrected for reddening using $E(B-V)=0.53$.
The vertical dotted lines represent the wavelengths $\lambda\lambda$
1512, 1878, and 2386 \AA~  at which the extinction law takes the same value.
With the adopted value of reddening, the stellar continuum underlying the
many emission lines is well represented by a straight line all over the full
spectral range. Saturated data points in
the emission lines are labeled with pluses.  
\label{UVreddening}}
\end{figure}

The Galactic extinction curve \citep{sea79} shows a pronounced broad maximum
around 2175 \AA~ due to dust absorption. Since  it takes the same value
$X(\lambda) = A(\lambda)/E(B-V) \approx 8$ at $\lambda = 1512$, 1878, and 2386
\AA, the slope of the straight line passing through the continuum
points at these wavelengths is insensitive to $E(B-V)$ in a $(\lambda , \log
F_\lambda)$ plot.  This circumstance can be used to get a reliable estimate
of $E(B-V)$ as that in which the stellar continuum becomes closely linear in
the 1512--2386 \AA~ region, and passes through the continuum points at the
above wavelengths.  From 8 pairs of short and long wavelength {\it IUE}
spectra taken from day 6 to day 69, i.e. during or close to the nebular phase, we
have in this way found $E(B-V) = 0.53 \pm 0.05$, a value that we adopt in the
following. Examples of {\it IUE} spectra of V838 Her corrected with $E(B-V) =
0.53$ are reported in Figure \ref{UVreddening} for day 10, 12, 15 and 22.

In principle, we could have estimated the $E(B-V)$ color excess also from the
observed relative intensities of the \ion{He}{2} 1640 \AA~ Balmer line, and
the 2734 and 3203 \AA~ Paschen recombination lines, compared with theoretical
ratios, as done, for example in the case of GQ Mus  (Paper III).
Such lines were however too faint to be measured accurately.


\subsection{Evolution of the UV Continuum}
\label{uv_evolution_cont}
\begin{figure}
\epsscale{1.1}
\plotone{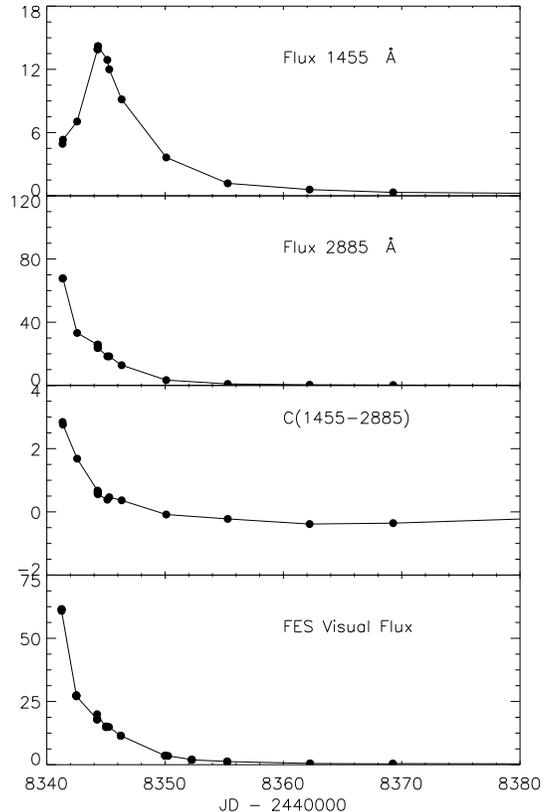}
\caption{Evolution of the continuum fluxes at 1455 \AA~ and 2885 \AA,
of the ultraviolet color index $C(1455 - 2885)$, and of the $V_{\rm FES}$ 
visual flux of V838 Her. Fluxes are in units of 10$^{-13}$ 
erg cm$^{-2}$ s$^{-1}$ \AA$^{-1}$, not corrected for reddening. 
Only the color index has been corrected for reddening using
$E(B-V) = 0.53$.
\label{UV1455evolution}}
\end{figure}

We have measured the mean flux in two narrow bands 20 \AA~ wide centered at
1455 \AA~ and 2855 \AA, selected to best represent the UV continuum because
little affected by emission lines \citep{cas02}. Figure \ref{UV1455evolution}
shows the time evolution of the $F(1455$~\AA) and $F(2885$ \AA) fluxes and of
the UV color index $C(1455-2885) = -2.5 \log[F(1455$~\AA)$/F(2885$~\AA)].  The
measurements were made on well exposed low resolution large aperture spectra.
Figure \ref{UV1455evolution} reports also, for comparison, the visual light
curve obtained from the Fine Error Sensor (FES) counts on board {\it IUE},
once corrected for the time dependent sensitivity degradation
\citep[see][for details on the FES calibration]{cas04a}.

It appears from Figure \ref{UV1455evolution} that flux maximum is reached
later in the UV 1455 \AA~ band than in the 2885 \AA~ and visual bands, as a
result of the progressive shift of the maximum emissivity towards 
shorter wavelengths.  The progressive hardening of the UV spectrum stops
after day 10, when the C(1455 - 2885) color index stabilizes around a value of
zero. Both these features are a common property of the novae studied by
\citet{cas02}.


\subsection{Evolution of the UV emission Lines}
\begin{figure}
\epsscale{1.1}
\plotone{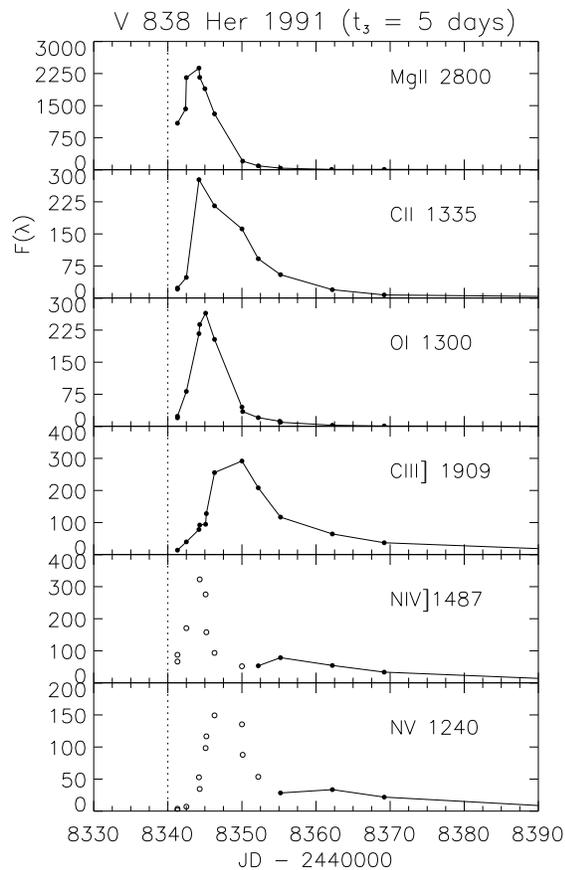}
\caption{Evolution of the observed fluxes in the most prominent emission lines
of V838 Her, in units of 10$^{-12}$ erg cm$^{-2}$ s$^{-1}$. Open dots
in the \ion{N}{4]} and \ion{N}{5} panels refer to features not due to
these transitions.}
\label{UVlinesevolution}
\end{figure}

\begin{figure}
\epsscale{1.0}
\plotone{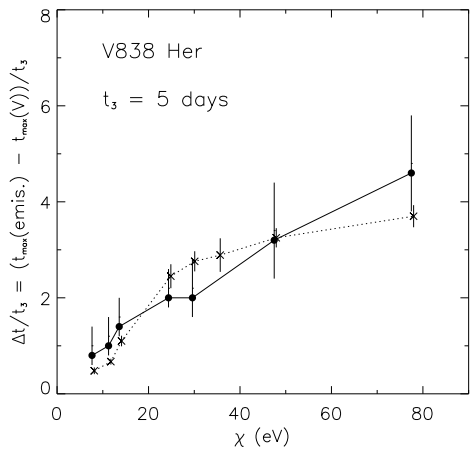}
\caption{
Delay of the flux maximum of the emission lines, normalized to the
$t_3$ time, as a function of the corresponding ionization potential for
V838 Her (filled circles). The emission lines considered, in order of 
increasing ionization potential, are: \ion{Mg}{2} 2800 \AA, \ion{C}{2} 
1335 \AA, \ion{O}{1} 1300 \AA, \ion{C}{3]} 1909 \AA, \ion{N}{3]} 1750 \AA,
\ion{N}{4]} 1487 \AA, and \ion{N}{5} 1240 \AA. We also show, for comparison,
the average values of CO novae (crosses) in \citet{cas05}. In this case, 
the x-axis has been shifted by 0.5 eV to make the error bars visible and 
the \ion{O}{3} 1660 \AA~ line is added between the \ion{N}{3]} 
and \ion{N}{4]} lines. 
}
\label{UVionization}
\end{figure}

Figure \ref{UVlinesevolution} reports, as a function of time, the observed
flux in the most prominent UV emission lines, as measured by us from the
available {\it IUE} low resolution spectra. The figure shows that the maximum
emission progressively shifts from low to high ionization lines, such
that \ion{Mg}{2} 2800 \AA~ is the first to reach maximum, followed by
\ion{C}{2} 1335 \AA, \ion{O}{1} 1300 \AA, \ion{C}{3]} 1909 \AA, \ion{N}{4]}
1487 \AA, and \ion{N}{5} 1240 \AA ~maximum.
The time delay of maximum emission $\Delta t$ with respect to the visual
maximum does in fact increase with increasing the line ionization potential
$\chi$ [eV], as better shown in Figure  \ref{UVionization} where $\Delta t$,
once normalized to the $t_3$ time, is plotted as a function of $\chi$
for the different emission lines.  Here, $t_3$ is the time in which the 
visual magnitude drops
by 3 magnitudes from the optical peak. Such a behavior is consistent with that
observed in a sample of seven CO novae, whose mean data from \citet{cas05}
are overplotted for comparison.  In particular, note that the maximum emission
in \ion{C}{3]} 1909 \AA~ and \ion{N}{3]} 1750 \AA~takes 
place after a time of $\Delta t/ t_3$ $\approx $
2, as in CO novae. This time has been identified in Cassatella et al. as the
start of the pre-nebular phase.


\subsection{High Resolution Emission Line Profiles}

\begin{figure}
\epsscale{1.0}
\plotone{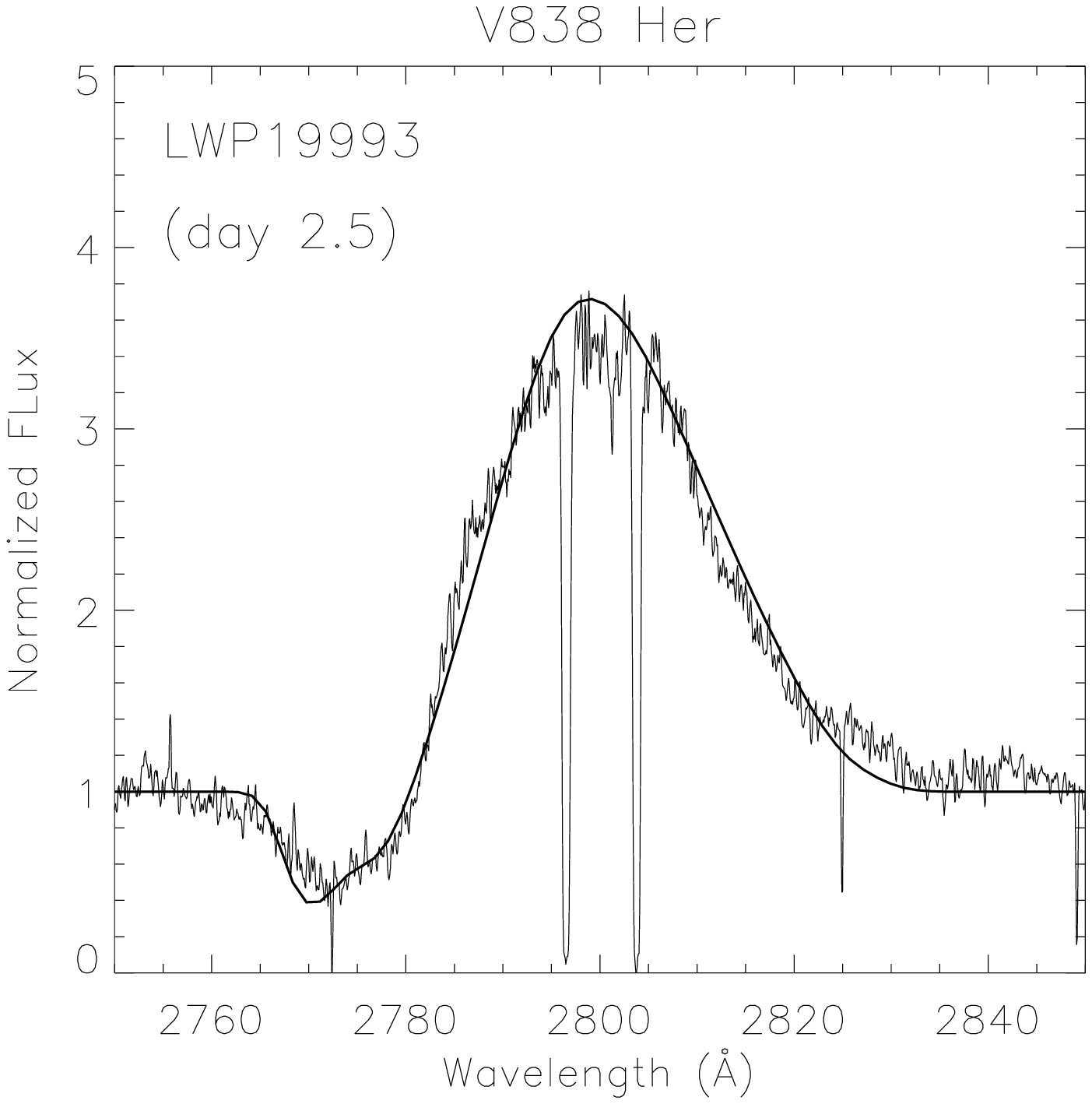} 
\caption{P Cygni profile of the \ion{Mg}{2} 2800 \AA~ doublet from the
$IUE$ spectrum LWP19993 of V838 Her taken on day 2.5.  The observed profiles
have been fitted with a theoretical P Cygni profile, obtained through
the SEI method \citep{lam87, gro89}, indicated as a thick line. The
derived wind terminal velocity is $\approx$ 3000 km~s$^{-1}$.  Fluxes are
normalized to the local continuum.  Note the presence of the narrow
interstellar components of the doublet at about the laboratory wavelength.}
\label{UVMg2}
\end{figure}

Good quality $IUE$ high resolution spectra of V838 Her are available only at
two dates: day 2.5 (LWP 19993) and day 15 (SWP41317). The most prominent
feature in the long wavelength spectrum of day 2.5 is the broad P Cygni
profile of the \ion{Mg}{2} 2800 \AA~ doublet (Figure \ref{UVMg2}). A comparison
with theoretical P Cygni profiles computed with the SEI (Soblolev Exact
Integration) method \citep{lam87,gro89} indicates that the terminal velocity
of the wind is $\approx$ 3000 km~s$^{-1}$. The huge emission compared with the
absorption component, usual in novae, is due to efficient population of the
upper levels of the doublet by electron collisions.

\begin{figure}
\epsscale{1.0}
\plotone{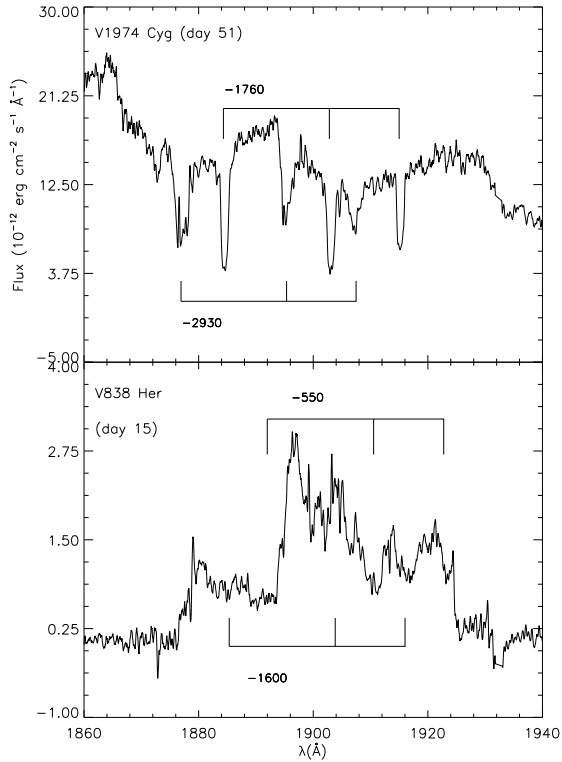}
\caption { Spectra of V838 Her and V1974 Cyg around 1900 \AA.  In the
case of V838 Her, a strong emission line from the \ion{C}{3]} 1907-1909
\AA~ doublet is present and appears eroded by overlying absorption from
the principal and diffuse system components of the \ion{Fe}{3} UV 34
multiplet. On day 15, these components are seen at $-550$ km~s$^{-1}$ and
$-1600$ km~s$^{-1}$, respectively. In V1974 Cyg, the \ion{C}{3]} doublet
emission is not present yet (it will appear at a later stage), but it is
easy to appreciate the presence of the principal and diffuse absorption
components of the \ion{Fe}{3} UV 34 multiplet at $-1760$ km~s$^{-1}$ and
$-2930$ km~s$^{-1}$.}
\label{UVcarbonio}
\end{figure}

Another interesting line profile is that of the \ion{C}{3]} 1906.68-1908.73
\AA~ doublet shown in Figure \ref{UVcarbonio}, which is the strongest emission
line observed in the short wavelength spectrum of day 15. As it appears
from the figure, the \ion{C}{3]} emission has a very complex profile.  One way
to interpret this profile is that the emitting region is very inhomogeneous in
terms of density and velocity structure. This possibility cannot be ruled out
a priori also in view of the presence of complex emission line profiles, e.g., in
the hydrogen Balmer lines \citep[see][]{iij09}.  The other possibility is that
the \ion{C}{3]} emission is eroded by overlying absorption from an outer
shell. A likely source of overlying absorption is the principal and diffuse
absorption components from the \ion{Fe}{3} UV 34 triplet at radial velocity of
$-500$ km~s$^{-1}$ and $- 1600$ km~s$^{-1}$, respectively.  Such components have
already been confidently identified to affect the \ion{C}{3]} emission line in
V1974 Cyg \citep{cas04b}.


\section{The model of Nova Light Curves}
\label{sec_model}

In this and previous papers, we have presented a unified model for the
IR, optical, UV, and supersoft X-ray light curves of several classical novae.
Our models are based on the optically thick wind theory of nova outbursts as
described in \citet{kat94h} and in Paper I.

\subsection{Optically Thick Wind Model}
After a thermonuclear runaway sets in, the photosphere of the white dwarf (WD) 
envelope expands greatly
to a giant size with $R_{\rm ph} \gtrsim 100 ~R_\odot$. The envelope settles
into steady-state around the optical peak.  The following decay phase can be
described by a sequence of steady state solutions \citep[e.g.,][]{kat94h}.
Using the same method and numerical techniques as in \citet{kat94h}, we have
followed the evolution of novae by connecting steady state solutions along the
decreasing envelope mass sequence.

The equations of motion, radiative diffusion, and conservation of energy, are
solved from the bottom of the hydrogen-rich envelope through the photosphere,
under the condition that the solution goes through a critical point of a
steady-state wind.  The winds are accelerated deep inside the photosphere so
that they are called ``optically thick winds.''  We have used updated OPAL
opacities \citep{igl96}.  Optically thick winds stop after a large part of 
the envelope is blown in the wind.
As one of the boundary conditions for our numerical
code, we assume that photons are emitted at the photosphere as a blackbody
with a photospheric temperature $T_{\rm ph}$.  

The UV 1455 \AA~ flux is estimated directly from blackbody emission instead of 
modeling WD atmosphere. This approximation is reasonably accurate in a very narrow 
wavelength range around UV 1455 \AA~ flux \citep[][and paper III]{cas02} near
its peak. Infrared and optical fluxes are
calculated from free-free emission by using physical values of our wind
solutions.  Physical properties of these wind solutions are
given in our previous papers \citep{hkn99,kat83,kat97,kat99,kat94h}.

We neglect helium ash layer which may develop underneath the hydrogen
burning zone in mass increasing WDs as in RS Oph \citep[see][]{hac07kl}. 
We suppose that the WD mass is decreasing in V838 Her, because 
metal enhancement is observed in V838 Her (see Table 
\ref{table_v838her_chemical_abundance}). 
Metal enrichment is an indicator of decreasing WD mass \citep[e.g.,][]{pri86,pri95}.

\subsection{Multiwavelength Light Curves}

In the optically thick wind model, a large part of the envelope is ejected
continuously for a relatively long period \citep[e.g.,][]{kat94h}.  After 
maximum expansion of the photosphere, the photospheric radius ($R_{\rm ph}$)
gradually decreases keeping the total luminosity ($L_{\rm ph}$) almost
constant.  The photospheric temperature ($T_{\rm ph}$) increases with time
because $L_{\rm ph} = 4 \pi R_{\rm ph}^2 \sigma T_{\rm ph}^4$.  The
maximum emission shifts from optical to supersoft X-ray through UV and 
extreme ultraviolet.  This causes the optical luminosity to
decrease and the UV luminosity to increase with a timescale that
depends strongly on the WD mass and weakly on the chemical composition of
the envelope \citep[e.g.][Paper I]{kat97}.

\subsection{Universal Decline Law of Free-Free Light Curves}
\label{subsec_universallaw}

We calculated the optical and IR fluxes of free-free emission outside the
photosphere of optically thick winds by 
\begin{equation}
F_\nu \propto \int N_e N_i d V
\propto \int_{R_{\rm ph}}^\infty {\dot M_{\rm wind}^2
\over {v_{\rm wind}^2 r^4}} r^2 dr
\propto {\dot M_{\rm wind}^2 \over {v_{\rm ph}^2 R_{\rm ph}}},
\label{free-free-wind}
\end{equation}
(see Paper I for more details), where $F_\nu$ is the flux at the frequency
$\nu$, $N_e$ and $N_i$ are the number densities of electrons and ions,
respectively, $V$ is the emitting volume, $R_{\rm ph}$ is the photospheric
radius, $\dot M_{\rm wind}$ is the wind mass loss rate, and $v_{\rm ph}$
is the velocity at the photosphere.
We also assume that $N_e \propto \rho_{\rm wind}$, $N_i \propto \rho_{\rm
  wind}$, and use the continuity equation, i.e., $\rho_{\rm wind} = \dot
M_{\rm wind}/ 4 \pi r^2 v_{\rm wind}$, where $\rho_{\rm wind}$ and $v_{\rm
  wind}$ are the density and velocity of the wind, respectively. Finally, we
assume that $v_{\rm wind}= {\rm const.}= v_{\rm ph}$ outside the photosphere.

This equation has successfully reproduced the optical and infrared light
curves of several novae in outburst \citep[see papers of this series;][]
{hac06a,hac09,kat08}. 
In the case of V838 Her, the early decay (until day 6) 
of $J$, $H$, and $K$-bands are almost parallel to each other as
shown later in Figure \ref{light}, whereas $J$ is not shown for clarity; 
This is one of the characteristic properties of free-free emission (see
Appendix of Paper III). Also the spectrum of V838 Her is consistent with that of 
free-free emission as mentioned in Section \ref{introduction}. Thus, we can 
safely apply our method to V838 Her.

After the optically thick wind stops, the total mass of the ejecta
remains constant with time.  The flux from such homologously expanding
ejecta is
\begin{eqnarray}
F_\nu \propto \int N_e N_i d V
\propto \rho^2 V \propto {M_{\rm ej}^2 \over V^2} V~
 \propto  R^{-3}   \propto  t^{-3},
\label{free-free-stop}
\end{eqnarray}
where $\rho$ is the mean density, and $M_{\rm ej}$ is the
ejecta mass.  We assume that the ejecta are expanding at a constant
velocity, $v$.  So we have $R = v t$, where $t$ is the time after the
outburst.  The proportionality constants in equations
(\ref{free-free-wind}) and (\ref{free-free-stop}) cannot be determined a
priori because radiative transfer is not calculated outside the photosphere; 
these were determined using the procedure described below.
$F_\lambda$ is also represented by equations (\ref{free-free-wind}) and 
(\ref{free-free-stop}), but the proportionality constant depends on 
wavelength  since  $F_{\lambda} \propto \lambda^{-2} F_{\nu}$.

If we assume an expanding shell with constant thickness, the flux is 
proportional to $F_\nu \propto t^{-2}$ as discussed in V723 Cas \citep{eva03}.
In V1500 Cyg and V1974 Cyg, however, we found that the flux decline rate can be well 
represented by $t^{-3}$ (see Paper I). As we will see later, the decline rate 
of V838 Her is consistent with  $t^{-3}$.

As the proportional constant in equation (\ref{free-free-stop})
 cannot be determined theoretically, and there is no representative timescale,
 this later part of the light curve does not contain any information on 
the WD mass. Therefore, we do not use this part in our light curve fitting.


\begin{figure}
\epsscale{1.1} 
\plotone{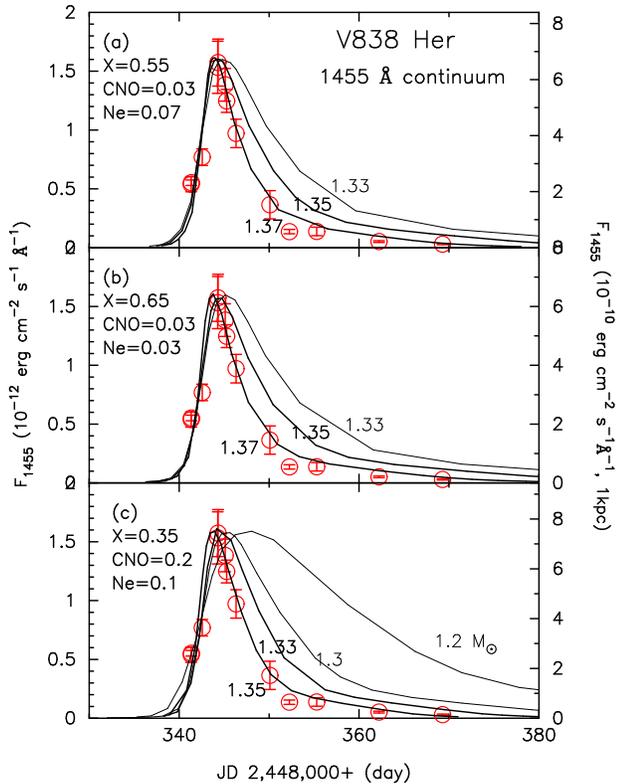}
\caption{UV 1455 \AA~ light curve-fitting for V838 Her. Large open circles 
with error bar: observed UV 1455 \AA~ flux and root mean square of errors
(units of the left-hand-side axis).
  Three different sets of chemical composition are assumed. 
  (a) $X=0.55,~X_{\rm CNO}=0.03$, $X_{\rm Ne}=0.07$, and $Z=0.02$, (b)
  $X=0.65,~X_{\rm CNO}=0.03$, $X_{\rm Ne}=0.03$, and $Z=0.02$ and
  (c)$X=0.35,~X_{\rm CNO}=0.2$, $X_{\rm Ne}=0.1$, and $Z=0.02$. 
  The WD mass is attached to each light curve.
  Theoretical flux $F_{1455}$ is presented
  for an arbitrarily assumed distance of 1.0 kpc and no absorption (the scale
  in the right-hand-side is for $1.37 M_{\odot}$ in (a) and (b), but for 
   $1.35 M_{\odot}$ in (c). For other
  curves scales are normalized arbitrarily).
\label{lightUV3}}
\end{figure}

\subsection{System Parameters of Optically Thick Wind Model}

The light curves of our optically thick wind model are parameterized by the WD
mass ($M_{\rm WD}$), chemical composition of the envelope ($X, Y$, $X_{\rm CNO}$,
$X_{\rm Ne}$, $Z$), and the envelope mass ($\Delta M_{\rm env, 0}$) at the
time of outburst (JD 2,448,340.0).
Note that the metal abundance, $Z$, includes
carbon, nitrogen, oxygen, and neon with solar composition ratios and that
$X_{\rm CNO}$ and $X_{\rm Ne}$ denote additional excesses.

Table \ref{table_v838her_chemical_abundance} summarizes the chemical
abundance determinations so far available for V838 Her, as first provided by
\citet{van96}, superseded by the later re-determinations by \citet{van97},
and extended by \citet{sch07} from spectra taken at four epochs (day 28, 60,
79 and 148).  We thus assume the chemical composition of V838 Her to be 
$X=0.55$, $Y=0.33$, $X_{\rm CNO}=0.03$, $X_{\rm Ne}=0.07$, and $Z=0.02$. 
We call it the ``standard'' composition set.  
We also assumed two other sets of composition for comparison. The first case is 
$X=0.65$, $Y=0.27$, $X_{\rm CNO}=0.03$, $X_{\rm Ne}=0.03$, and $Z=0.02$,  
expressing less enrichment of heavy elements such as in 
V382 Vel and QU Vul, and the second case is 
$X=0.35$, $Y=0.33$, $X_{\rm CNO}=0.2$, $X_{\rm Ne}=0.1$, and $Z=0.02$
which may be an extreme case of metal enrichment for V838 Her 
but appropriate composition for V1974 Cyg and V351 Pup, 
although abundance estimates are very different among different authors
(see Table 1 in Paper I which lists observational estimates of abundance 
of nova ejecta).

The latter two recent determinations both indicate an excess
of helium with respect to hydrogen, i.e., $X/Y=1.9$ \citep{van97} and 1.8
\citep{sch07}, against the solar value of 0.7/0.28 = 2.58.  In massive WDs
about one tenth of hydrogen is consumed by nuclear burning to produce thermal
and gravitational energy during the early phase of the outburst
\citep{pol95,pri95}.  Supposing that $\Delta X=0.1$ of hydrogen 
is converted into helium, the hydrogen/helium ratio becomes $X/Y=$ 
$(0.7-0.1)/(0.28+0.1)=1.6$.  If, in addition, a fraction $10 \%$ of the
envelope mass is dredged up and mixed into the envelope, the above envelope
composition is further changed to $X=0.545, Y=0.345 $, and $X_{\rm
CNO}+X_{\rm Ne}+Z=0.11$.  This is very consistent with the observational
results by \citet{van97} and \citet{sch07} in Table
\ref{table_v838her_chemical_abundance}.

\section{Light Curve Fitting}
\label{sec_lightcurvefitting}

We have searched for the model light curves that best fit the observational data
among models with WD mass 1.2, 1.3, 1.33, 1.35, and $1.37~M_\odot$, for the three sets of
chemical abundance quoted above.
As shown in previous papers of this series, we have proven the light 
curve fitting method to be a powerful tool to obtain the intrinsic nova 
parameters. In particular, we have shown that the theoretical
light curves of UV 1455 \AA~ continuum are very sensitive to the WD
mass, so that the WD mass can be determined with a good accuracy.

Figure \ref{lightUV3} shows the fits to the UV 1455\AA~ light curve of V838
Her for three different chemical composition and various WD masses.  
Figure \ref{lightUV3}a) shows that $M_{\rm WD}= 1.37~
M_\odot$ model is in a good agreement with the observed data but models of
$M_{\rm WD} \leq 1.33~ M_{\odot}$ produce too slow declines.  Models of 
1.2 and 1.3 $M_{\odot}$ are omitted because they are too slow to be 
compatible with the observation. Similar results are 
obtained if we increase the hydrogen content and decrease the heavy elements 
slightly to $(X,
X_{\rm CNO},X_{\rm Ne})$ =$(0.65,0.03,0.03$) as in Figure \ref{lightUV3}b. 
On the other hand, if we decrease the hydrogen and increase the heavy elements 
to $X=0.35, X_{\rm CNO}=0.2$, and $X_{\rm Ne}=0.1$ (Figure \ref{lightUV3}c), which 
may be an extreme case for V838 Her, the $1.35 M_\odot$ model 
shows a good fitting and less massive WDs ($M_{\rm WD} \leq 1.3$) are probably 
excluded.  Therefore, from the UV light curve fitting, we see that the 
WD in V838 Her is as massive as $M_{\rm WD} \geq 1.35~M_\odot$. 


\begin{figure}
\epsscale{1.1}
\plotone{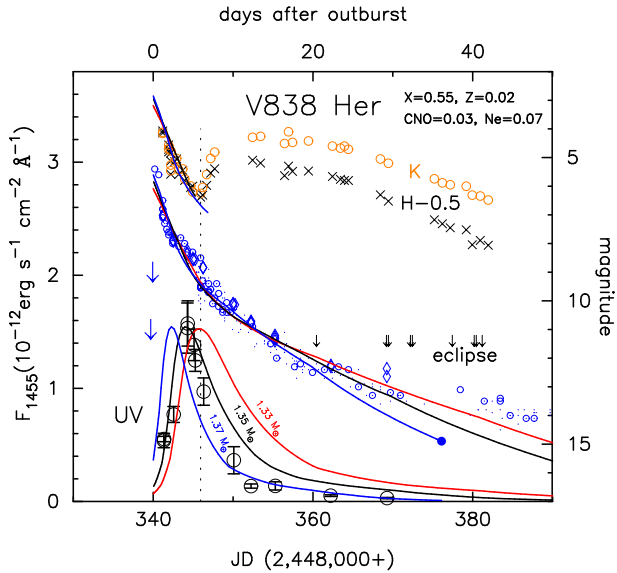}
\caption{ Theoretical and observed optical, infrared, and UV light curves
of V838 Her.  Theoretical light curves refer to models with $M_{\rm
WD}=1.37~M_{\odot}$ (blue), 1.35 ~$M_{\odot}$ (black), and $1.33~
M_{\odot}$ (red).
Observed visual and V magnitudes are denoted by small open circles (IAU
Circulars 5222, 5223, 5224, 5226, 5228, 5229, 5233, 5234, 5238, 5254 and
5265), and by diamonds ({\it IUE} $V_{\rm FES}$: \citet{cas04a}). Two large 
downward arrows
indicate optical upper limit observations from IAU Circular 5265.
Small downward arrows denote the dates of eclipse minima reported by
\citet{lei92} and by \citet{tkat91}. Infrared data are taken
from IAU Circulars 5223, 5224, 5228, 5230, 5232, 5236, 5246, 5254 and from
\citet{har94}. $K$ and $H$ magnitudes are denoted by middle-sized open circles
and crosses, respectively. $H$ magnitudes are shifted up by 0.5 mag.
Large open circles denote the observed UV 1455~\AA~ band flux
\citep[][except one newly obtained point]{cas02} in units of 
$10^{-12}$~erg~s$^{-1}$cm$^{-2}$~\AA$^{-1}$. For theoretical UV curves, we assume 
the distance and absorption in Table \ref{table_model} to fit the observational data.
Vertical dotted line indicates 
the epoch of the IR minimum. 
\label{light}}
\end{figure}

Figure \ref{light} reports the model fits to the visual and infrared
light curves for the standard composition set. We have assumed that, 
at these wavelengths, the light
curve is dominated by free-free emission (i.e., can be calculated from
eq. [\ref{free-free-wind}]). Since its shape does not depend on
wavelength, one can apply the same light curve model to the optical and IR
data. Figure \ref{light} shows indeed that, if suitably upward-shifted, the
optical light curve also fits the infrared fluxes, at least until day 6,
when an IR flux minimum was reached (indicated as a vertical dashed line in
Figure \ref{light}), owing to thermal emission from dust (see Section \ref{introduction} and
Section \ref{sec_discussion}).  

Figure \ref{lightlog} provides, in logarithmic time, another view of
the model fit to the data.
This figure shows that, at very late phases, optical and
infrared fluxes decay along with the $F_\lambda \propto t^{-3}$ line as
predicted from equation (\ref{free-free-stop}). This circumstance does not
bring, however, any information on the WD mass.
As shown in Figures \ref{light} and \ref{lightlog}, among models with the WD mass
of 1.33 , 1.35 and 1.37~$M_\odot$, the 1.35 ~$M_\odot$ WD provides a best 
fit representation of the UV light curves as well as the optical and infrared 
light curves.  


\begin{figure}
\epsscale{1.1} 
\plotone{f9.epsi}
\caption{ Same as Figure \ref{light}, but for logarithmic time.
The AAVSO data (small dots) are plotted only until day 68, 
because after that time the nova becomes fainter than 14 mag, i.e., 
the brightness of a closeby star.  The
optical data for the late  phases (filled squares) are taken from
\citet{ing92} and \citet{szk94i}.  The IR data denoted by the asterisk
are taken from \citet{har94}. The emergence time of companion for the 1.35~$M_{\odot}$
WD model (day 9) and the first eclipse observation (day 21) are also
indicated by the small and large downward arrows.
\label{lightlog}}
\end{figure} 

\begin{figure}
\epsscale{1.1} 
\plotone{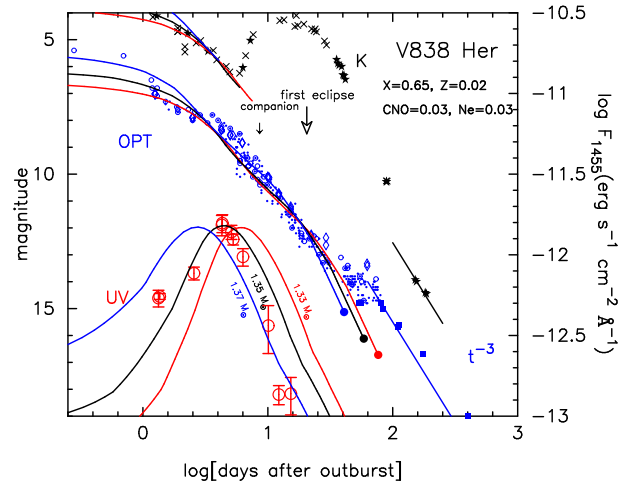}
\caption{Same as Figure \ref{lightlog} but for theoretical model with the 
composition set of less enhancement of heavy elements, i.e., $X=0.65,~X_{\rm CNO}=
0.03$, and $X_{\rm Ne}=0.03$.
\label{lightlog.compositionb}}
\end{figure}

\begin{figure}
\epsscale{1.1} 
\plotone{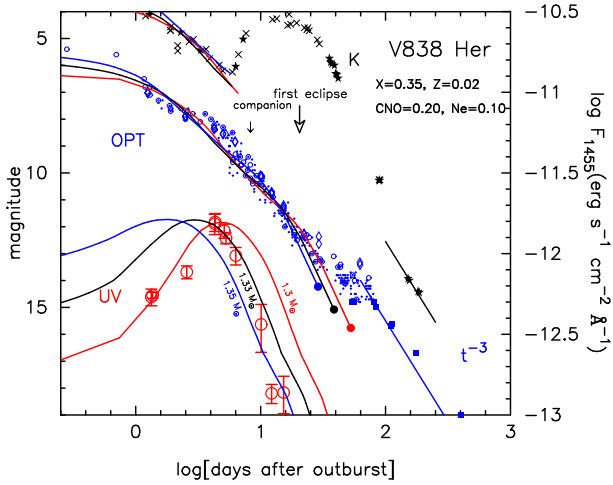}
\caption{Same as Figure \ref{lightlog} but for theoretical model with the
composition set of much enhancement of heavy elements, i.e., $X=0.35,~X_{\rm CNO}=
0.2$, and $X_{\rm Ne}=0.1$.
\label{lightlog.compositionc}}
\end{figure}

Figures \ref{lightlog.compositionb} and \ref{lightlog.compositionc} show the model fits
for the other two sets of chemical composition. 
In Figure \ref{lightlog.compositionb}, we get the best fit models of 
1.35 ~$M_\odot$ WD from fitting optical, IR and UV fluxes. In Figure 
\ref{lightlog.compositionc}, however, we cannot 
fit both of the UV and optical fluxes simultaneously. When we fit the optical, 
the model UV flux rises too earlier, and then we cannot obtain a good fit in the UV. 
This indicates that the assumed composition, i.e, $X=0.35,  X_{\rm CNO}=0.2$, and 
$X_{\rm Ne}=0.1$ is too metal rich and not appropriate for V838 Her.

To summarize we may conclude that the WD mass is $M_{\rm WD} = 1.35
\pm 0.02~M_\odot$.  The model parameters and the main results are summarized in
Table \ref{table_model}.


\begin{deluxetable*}{llllllll}
\tabletypesize{\scriptsize}
\tablecaption{Summary of the present model
\label{table_model}}
\tablewidth{0pt}
\tablehead{
\colhead{subject} &
\colhead{} &
\colhead{units}&
\colhead{model 1} &
\colhead{model 2} &
\colhead{model 3} &
\colhead{model 4} &
\colhead{model 5} \\
\colhead{} &
\colhead{} &
\colhead{} &
\colhead{1.37 $M_\odot$} &
\colhead{1.35 $M_\odot$} &
\colhead{1.33 $M_\odot$} &
\colhead{1.35 $M_\odot$} &
\colhead{1.33 $M_\odot$} 
}
\startdata
outburst day ($t_0$)& ...   && JD 2,448,340 &$\leftarrow$  &$\leftarrow$&$\leftarrow$ &$\leftarrow$ \\
$E(B-V)$ & ... & & 0.53  &$\leftarrow$  & $\leftarrow$& $\leftarrow$& $\leftarrow$ \\
$X$ & ... & & 0.55  &$\leftarrow$  &$\leftarrow$ & 0.65 & 0.35 \\
$Y$ & ... & & 0.33  &$\leftarrow$  &$\leftarrow$ &0.27  & 0.33 \\
$CNO$ & ...  && 0.03  &$\leftarrow$  &$\leftarrow$ & 0.03 & 0.2\\
$Ne$ & ... & & 0.07  &$\leftarrow$  &$\leftarrow$ & 0.03 & 0.1 \\
$Z$ & ... &  &0.02  &$\leftarrow$   &$\leftarrow$ &$\leftarrow$ &$\leftarrow$ \\
distance from UV fit&... & kpc& 2.7 & $2.7 $& 2.6 & 2.6 & 2.8  \\
$M_{V, {\rm max}}$ from UV fit & ...  & mag& $-8.41$&$-8.40$ &$-8.32$&$-8.32$   &$-8.48$   \\
WD envelope mass & ...&$M_\odot$  & $2.5 \times 10^{-6}$ &  $2.9 \times 10^{-6}$& $3.5 \times 10^{-6}$  & $2.6 \times 10^{-6}$& $2.9 \times 10^{-6}$\\
mass lost by winds & ... &$M_\odot$ & $2.3 \times 10^{-6}$ &$2.6 \times 10^{-6}$& $3.0 \times 10^{-6}$ & $2.2 \times 10^{-6}$ & $2.5 \times 10^{-6}$\\
wind phase & ...  & days& 37 &52&68 & 58 & 39 \\
secondary mass&...& $M_\odot$ &0.76\tablenotemark{a} &$\leftarrow$ &$\leftarrow$ &$\leftarrow$ &$\leftarrow$ \\
separation & ... & $R_\odot$& 2.4 &2.4&2.4&2.4&2.4 \\
$R_1^*$ & ... & $R_\odot$& 1.0 &1.0 &1.0&1.0&1.0 \\
$R_2^*$ & ... & $R_\odot$& 0.8 &0.8 &0.8&0.8&0.8 \\
companion's emergence & ...&days & 10  &15 & 19  & 15 & 14
\enddata
\tablenotetext{a}{estimated from eq. (\ref{warner_mass_formula}).}
\end{deluxetable*}

\section{Discussion}
\label{sec_discussion}

\subsection{``Optical Drop'' in \citet{woo92}}\label{optical_drop}

\begin{figure}
  \epsscale{0.95} 
\plotone{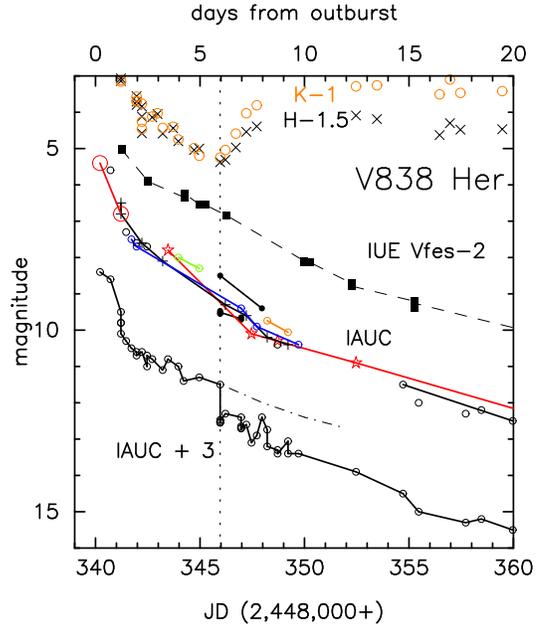} 
\caption{Optical and infrared light curves. The optical data are shown separately 
to assess their homogeneity.  From bottom to top, IAUC+3: data from IAU Circulars, 
shifted by 3 mag downward, in which all the data are connected along the reported 
observing time. IAUC: same data of IAU Circulars but are connected between points only 
obtained by the same observer. $IUE$ $ V_{\rm FES}-2$: $IUE$ $ V_{\rm FES}$ data 
connected 
by a dashed line (shifted 2 mag upward). $K-1$ and $H-1.5$: infrared $K$ and $H$ data,
shifted by 1.0 and 1.5 mag upward, respectively.  The time of the ``drop in
optical magnitude'' proposed by \citet{woo92} is indicated with a vertical
dashed line, and their ``unabsorbed light curve'' with a dot-dashed line.
\label{light.iaucconnect}} 
\end{figure}

\citet{woo92} and \citet{har94} claimed that the rise of infrared fluxes
starting on JD 2,448,346 (day 6) was accompanied by a sudden drop of the
visual magnitude by $\Delta V \sim 0.9$.  
\citet{woo92} proposed that the sudden drop of the visual flux was caused by
an optically thick clump of matter ejected toward the earth on day 6
(indicated by the vertical dotted line in Figure \ref{light.iaucconnect}).
Afterwards, the optical flux continued to decline but at a substantially lower
level than extrapolated from the light curve of previous days, shown by 
a dash-dotted line in Figure \ref{light.iaucconnect} \citep[which mimics the dashed
curve in Figure  1 of][]{woo92}.


The visual magnitudes used by \citet{woo92} and by \citet{har94} were taken from
IAU Circulars.  These data were, however, obtained by 
different observers. We have tested self-consistency of the different sets
by connecting the data obtained by the same observer (Figure 
\ref{light.iaucconnect}). It clearly appears from the figure that there
are systematic differences between the data sets from different observers,
and also that no observer reported the sudden drop at JD 2,448,346, 
in agreement with the smooth decline shown by the homogeneous set of
the $IUE ~V_{\rm FES}$ magnitudes, also given in the same figure. Moreover, 
the magnitudes corresponding to just before and after the sudden drop are 
8.5 mag at Mar. 30.46 UT and 9.54 mag at Mar. 30.466 UT, respectively, which 
are obtained by different observers. 
If we take the reported observation time as a three-digit number, the magnitude 
should dropped by 0.006 day = 8.6 minutes. 
This is too short to have some significant change in stellar luminosity. 
Therefore, we conclude
that the claimed $V$ magnitude drop is an artifact caused by connecting 
the different data set with systematic difference.

Should the 0.9 mag optical drop be real and due to dust absorption from
a clump of matter ejected along the line of sight, one would expect to
observe a simultaneous strong absorption of the UV flux.  The decrease in the
UV flux is estimated to be a factor of $10^{(8.3/3.1)*0.9}=257$. Figure
\ref{light} shows, however, no indication of such a drastic decrease in the UV
flux.

\subsection{Dust Formation: Comparison with V1668 Cyg}
\label{sec:dust}
After the local minimum on day 6, the IR fluxes increased to reach a local
maximum around day 12 (in $H$ band), being ascribed to thermal emission from an
optically thin dust shell \citep{cha92,woo92,kid93,har94}. 
As shown in Figure \ref{light}, however, there is no
clear evidence for a corresponding decrease in the UV fluxes. In this
subsection we make a comparison with nova V1668 Cyg in order to understand to
what extent dust absorption can cause significant effects on the UV and
optical fluxes in the case of V838 Her.

V1668 Cyg is a well studied moderately fast nova. It developed an
optically thin dust shell giving rise to the infrared second maximum 
$\sim$ 60 days after the optical peak \citep{sti81}. In
correspondence to the second IR maximum, the UV 1455 \AA~ flux suddenly
dropped by $1/2 \sim 1/3$ of what expected from a linear decline law
\citep{kat07h}. The evolution of V1668 Cyg ($t_3$ $\sim$ 23 days) was much
slower than that of V838 Her, due to its smaller WD mass ($\sim 0.95~M_\odot$,
see Paper I).  Consequently, the ejected mass of V1668 Cyg is expected to be
much larger than that of V838 Her. Namely, since the ejected mass of V1668 Cyg 
is estimated to be  $5.8 \times 10^{-5}~M_\odot$ \citep{kat07h}, 
while we estimate that it was $2-3 \times
10^{-6}~M_\odot$ for V838 Her from our best fit models (see Table \ref{table_model}). 
One should thus expect that the dust mass of V838 Her was roughly ten 
times smaller.

Observationally, \citet{geh80} estimated for V1668 Cyg the mass of dust
 (condensed carbon) to be $2.5 \times 10^{-8}~M_\odot$ and shell mass of hydrogen 
to be $2 \times 10^{-5}~M_\odot$, which is consistent with the above value of 
total ejecta mass $5.8 \times 10^{-5}~M_\odot$. 

Also, \citet{cha92} estimated the total grain mass of V838 Her to be
$1.6\times 10^{-8}~M_\odot$ assuming distance of $d=8.3$ kpc.  As the mass
estimate is proportional to $d^{2}$, where $d$ is the distance 
to the star, this corresponds to $1.7 \times 10^{-9}~M_\odot$
for $d=2.7$ kpc here adopted (see Section 5.3 below).  \citet{har94}
estimated the dust mass of V838 Her to be $1.1 \times 10^{-8}~M_\odot$ at dust shell
maximum for $d=5.5$ kpc, which is proportional to the IR luminosity, therefore, 
a smaller value $2.7 \times 10^{-9}~M_\odot$ is derived for $d=2.7$ kpc.

In any case, the dust mass of V838 Her appears to be one tenth 
 of dust mass of V1668 Cyg, a value that is not expected to affect 
significantly
the UV fluxes even at the time of the second infrared maximum, explaining then
the absence of any quick drop in the UV 1455 \AA~ light curve.

\subsection{Distance}
\label{sec_distance}

The interstellar extinction and the distance were estimated with various
methods \citep[][for summary]{har94,van96}, the values of which are 
largely scattered as $E(B-V)=0.3-0.6$ and $d=2.8-10$ kpc (see Table 
\ref{table_observation}).  One
of the typical way to estimate the distance is to use $t_3$ time or using
statistical properties such as a given magnitude at a certain stage of
light curve. 

\begin{figure} 
\epsscale{1.0} 
\plotone{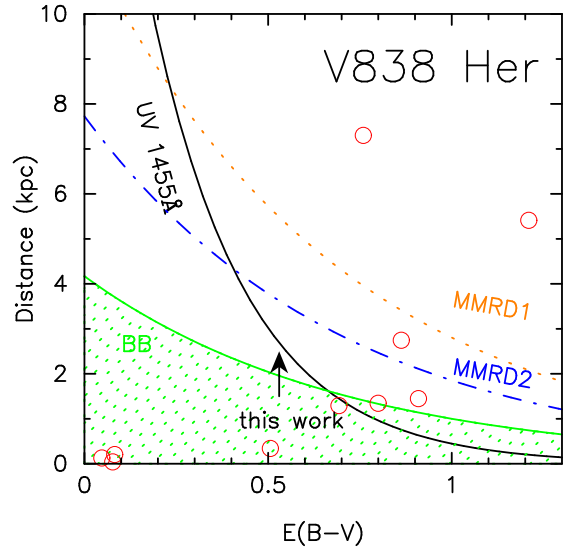}
\caption{Distance-reddening relations for V838 Her. Solid line
labeled ``UV 1455 \AA'' (see Eq. [\ref{eq:uvdis}]), corresponds to
our best fit model $M_{\rm WD}= 1.35 ~M_\odot$. Dotted curve represents 
the maximum magnitude vs. rate of decline relation 
calculated from Schmidt-Kaler's law, with $t_3 = 5.3$ days  
(labeled ``MMRD1'', eq.  [\ref{modulus-mmrd}] ) 
; the dash-dotted curve shows another MMRD relation, by \citet{del95}, 
 with $t_2 = 1.4$ day (labeled "MMRD2", eq. [\ref{modulus-mmrd2}]). 
``BB'' is the line of eq.(\ref{BBdistance}).
 An arrow indicates our reddening
determination $E(B-V)= 0.53$.  Open circles denote nearby stars' distance vs.
$E(B-V)$ (taken from (46/2) in \citet{nec80}).  
See text for more details.  
\label{AvD}}
\end{figure}

Let us begin with the distance estimate using the Maximum Magnitude 
versus Rate of Decline (MMRD) law. The absolute $V$ magnitude at maximum
$M_{\rm V, max}$ can be estimated from the $t_3$ time through Schmidt--Kaler's MMRD
relation \citep{sch57}:
\begin{equation}
M_{\rm V,max} = -11.75 + 2.5 \log t_3.
\label{schmidt-kaler-law}
\end{equation}
Then the distance is obtained from 
\begin{equation}
(m- M)_{\rm V,max} = -5 ~+~ 5 \log~d ~+~ 3.1 E(B-V),
\label{modulus}
\end{equation}
if  $E(B-V)$ is known.  If we adopt $t_3=5.3$ from the optical maximum of
$m_{\rm V}=5.4$ \citep{har94}, this equation gives $M_V=-9.94$.  With the apparent
visual magnitude of $m_{\rm V}=5.4$ and $E(B-V)=0.53$, we get $d=5.49$ kpc. If
$E(B-V)$ is not fixed, we have
\begin{equation}
(m- M)_{\rm V,max} = -5 ~+~ 5 \log~d ~+~ 3.1 E(B-V)= 15.34,
\label{modulus-mmrd}
\end{equation}
which is plotted in Figure \ref{AvD} (labeled ``MMRD1''). 


The Schmidt--Kaler law (Eq. (\ref{schmidt-kaler-law})) yields a bit 
larger magnitudes 
for a faster nova. \citet{del95} derived the following MMRD relation 
which is a good representation of fast novae,
\begin{equation}
M_{\rm V,max} = -7.92 ~-~ 0.81 \arctan~{{1.32-\log(t_2)}\over0.23}.
\label{dellavalle.MMRD}
\end{equation}
For $t_2=1.4$ \citep{har94} this equation gives $M_V=-9.04$. 
With the apparent visual magnitude of $m_{\rm V}=5.4$ and $E(B-V)=0.53$, 
we get $d=3.6$ kpc. If $E(B-V)$ is not fixed, we have
\begin{equation}
(m- M)_{\rm V,max} = -5 ~+~ 5 \log~d ~+~ 3.1 E(B-V)= 14.44,
\label{modulus-mmrd2}
\end{equation}
which is plotted in Figure \ref{AvD} (labeled ``MMRD2'').

Another way to estimate the distance to V838 Her comes from comparing the
observed light curve of 1455 \AA~ band with the corresponding model fluxes
\citep[see Paper I, Paper III, and][]{kat05h, kat07h}.  We show that 
the calculated flux at $\lambda$ = 1455 \AA~ at
a distance of 1~kpc for Model 2 is $F_{\lambda}^{\rm mod}$ =
6.6 $\times 10^{-10}$~erg~cm$^{-2}$~s$^{-1}$~\AA$^{-1}$ at the UV peak 
in Figure \ref{lightUV3}a. The
corresponding observed peak flux is $F_{\lambda}^{\rm obs}$=1.57 $\times
10^{-12}$~erg~cm$^{-2}$~s$^{-1}$ ~\AA$^{-1}$.  From these values one obtains
the following relation between distance and reddening:
\begin{equation}
  m_{\lambda}^{\rm obs} - M_{\lambda}^{\rm mod}=
5 \log ({d \over {1\mbox{~kpc}}}) + A_{\lambda} E(B-V) = 6.56
\label{eq:uvdis}
\end{equation}
where $m_{\lambda} = -2.5 \log(F_{\lambda})$, and $A_{\lambda}= 8.3$ at
$\lambda=1455$ \AA~ \citep{sea79}. This curve is reported in Figure \ref{AvD},
and is labeled ``UV 1455\AA''.   Equation (\ref{eq:uvdis})
provides a distance of $2.7 \pm 0.5 $~kpc for $E(B-V)=0.53 \pm 0.05$.
If we take the upper and lower values of the UV peak 
$(1.57 \pm 0.20) \times 10^{-12}$ ~erg~cm$^{-2}$~s$^{-1}$ ~\AA$^{-1}$, 
we get the distance of 2.9 kpc and 2.5 kpc, respectively. This is 
however, an extreme case, because the fitting becomes much worse in Figure 
\ref{lightUV3}.
Therefore, we estimate the distance to be $2.7 \pm 0.5 $~kpc for $E(B-V)=0.53
 \pm 0.05$ from UV light curve fitting. 
Note that the distance hardly changes for all the models in Table 
\ref{table_model}.

One may estimate a lower limit for the distance assuming that the flux of 
free-free emission is larger than the blackbody flux in optical wavelength region.
As the optical free-free flux is dominated by photons reprocessed from shorter 
wavelength photons of blackbody continuum, free-free flux 
is larger than that of corresponding blackbody emission at the optical wavelength 
during the early phase of nova outburst \citep{hau97}.
Our experience of light curve fitting also suggest that this condition is 
empirically satisfied in several novae \citep[see][for V1974 Cyg]{kat07h}. 
Thus, we assume this condition, which is written as
\begin{equation}
m_{\rm V}- M_{\rm V} = 5  \log ({d \over {\mbox{~10 pc}}})~+~ 3.1 E(B-V) > 13.2. 
\label{BBdistance}
\end{equation}
The shadowed region labeled ``BB'' in Figure \ref{AvD} is where Equation
(\ref{BBdistance}) is not satisfied. The border gives the lower limit
for the distance to V838 Her, which is $d > 2.46$~kpc for the particular value
of $E(B-V)= 0.53$.

As shown in Table \ref{table_observation}, the distances in literature are very
scattered. The distance obtained from MMRD tends to be larger than ours as shown
in Figure \ref{AvD}.  \citet{sta92} obtained a distance of $\sim 3.4$ kpc
and $E(B-V) \sim$ 0.6 from the comparison of total UV flux (LWP/SPW) including
both of line and continuum emission with that of nova LMC 1991 showing a similar
spectrum development.  This is similar but contrasted to our method in the
sense that we use only continuum emission around 1455 \AA~ and compare it with
the theoretical value.  Their value $\sim 3.4$ kpc is larger than ours, but
closer than other estimates using the MMRD relation.

\subsection{Emergence of the Companion}
 
In the early phase of the outburst, the photosphere of the WD envelope 
extends beyond the size of
the binary orbit. The companion is deeply embedded inside the WD photosphere. 
After maximum expansion, the photospheric radius shrinks owing to
wind mass loss. We can estimate the epoch when the companion appears
from the photosphere.

From Warner's (1995) empirical formula the mass of the donor star 
(companion) is expressed as:
\begin{equation}
{{M_2} \over {M_\odot}} \approx 0.065 \left({{P_{\rm orb}}
\over {\rm hours}}\right)^{5/4},
\mbox{~for~} 1.3 < {{P_{\rm orb}} \over {\rm hours}} < 9
\label{warner_mass_formula}
\end{equation}
which gives $M_2 = 0.76 ~M_\odot$ for $P_{\rm orb}=0.2976$ days (7.14
hr), in good agreement with the value $M_2 = 0.73 - 0.75~M_\odot$ obtained
by \citet{szk94i}, although they could not well constrain the mass of the 
primary component ($M_{\rm WD} > 0.62~M_\odot$).

The binary separation is obtained from Kepler's law to be $a = 2.41 ~R_\odot$
for $M_{\rm WD}= 1.35 ~M_\odot$. The effective radius of the Roche lobe is
$R_1^* = 1.0 ~R_\odot$ for the primary component (WD), and $R_2^* = 0.80
~R_\odot$ for the secondary. Here we use an empirical formula of $R_1^*$ and 
$R_2^*$ given by 
\citet{egg83}. In our model, the companion emerges from the WD envelope when
the photospheric radius of the WD shrinks to $R_{\rm ph} \sim 3.2 ~R_\odot$
(the separation plus $R_2^*$), $\sim 2.4 ~R_\odot$ (the separation), or 
$1.6~R_\odot$ (the separation minus $R_2^*$).  This occurs on day 7,
9 and 11, respectively, in our Model 2. The
emergence time of the companion (the central time) is shown in 
Table \ref{table_model} and also indicated by a small arrow
in Figures \ref{lightlog}, \ref{lightlog.compositionb} and \ref{lightlog.compositionc}.
As the eclipse is due to the occultation of
the WD, or of the central part of the accretion disk by the companion,
the eclipse must occur after the above three epochs of emergence.
Our estimated time of companion emergence is a rough estimate, but it is well 
before the first eclipse observation (day 21), so consistent with observation.

\subsection{Mass determination from the duration of the 1455 \AA~ maximum}

The duration of a UV nova outburst, as measured by the
full width at half maximum $t_{\rm FWHM}$ of the 1455 \AA~ light curve 
\citep{cas02} is tightly correlated with the nova decay
time $t_3$ which, in turn, is a primary indicator of the WD mass (Livio
1992). Detailed model calculations by \citet{hac06a} have confirmed the
primary dependence of $t_{\rm FWHM}$ on $M_{\rm WD}$ and have shown, in
addition, that there is secondary, weaker, dependence on chemical
composition. These results imply that the duration of a UV nova outburst can be
used as a useful tool to estimate the WD mass in nova systems.
 
\begin{figure} 
\epsscale{1.1} 
\plotone{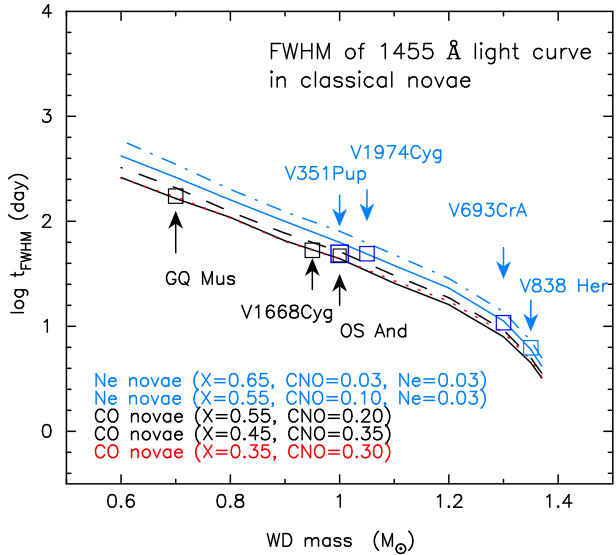}
\caption{Duration (FWHM) of UV nova outburst at 1455 \AA~ against the 
WD mass.  
{\it Dash-dotted line}: Neon novae ($X=0.65,~ Y=0.27,~ X_{\rm
CNO}=0.03,~ X_{\rm Ne}=0.03$).  {\it Top solid line}: Neon novae
($X=0.55,~ Y=0.30,~ X_{\rm CNO}=0.10,~ X_{\rm Ne}=0.03$).  {\it Dashed
line}: CO novae ($X=0.55,~ Y=0.23,~ X_{\rm CNO}=0.20$).  {\it Bottom solid
line}: CO novae ($X=0.45,~ Y=0.18,~ X_{\rm CNO}=0.35$).  {\it Dotted
line}: CO novae ($X=0.35,~ Y=0.33,~ X_{\rm CNO}=0.30$) (almost overlapped
with the lower solid line). We assume $Z=0.02$ for all the models.  Arrows
indicate the WD mass of individual novae. For V838 Her we adopt Model 2.  
For GQ Mus, we used the results in Paper III.
For V1668 Cyg and V1974 Cyg, the results in
Paper I. For V351 Pup, OS And and V693 CrA, the results in
\citet{kat07h}. 
Novae named in the upper side of the lines are Ne novae
whereas those in the lower side are CO novae.
\label{UVwidth}}
\end{figure}

In Figure \ref{UVwidth} we compare the theoretical relation of $t_{\rm FWHM}$
vs.  WD mass \citep{hac06a} with those obtained from the measured values of
$t_{\rm FWHM}$ \citep{cas02} and from the WD masses of seven classical novae,
including V838 Her, estimated from the previous works  \citep[namely from Paper I 
for V1668 Cyg and V1974 Cyg; from][]{kat07h} for V351 Pup, OS And and
V693 CrA, and from this paper for V838 Her). We recall that the masses of 
these novae were determined by using the light-curve fitting method not only 
on the UV 1455 \AA~ light curve, but also on the optical, infrared, and 
supersoft X-ray light curves. For GQ Mus the UV data are very scattered so we plot 
here the theoretically fitted value in Paper III.

The largest possible error in this mass determination arises from ambiguity of 
chemical composition because observationally determined values are sometimes very 
scattered in literature. If we adopt different chemical composition, we get 
different  $M_{\rm WD}$ even for the same  $t_{\rm FWHM}$ as shown in Figure \ref{UVwidth}.
In the case of V838 Her, however, the difference of mass determination 
 $\Delta M_{\rm WD}$ is relatively small, because in  very massive WDs,
 $t_{\rm FWHM}$ quickly decreases and then strongly depends on the WD mass 
but weakly on the composition. 


\begin{figure}
\epsscale{1.0}
\plotone{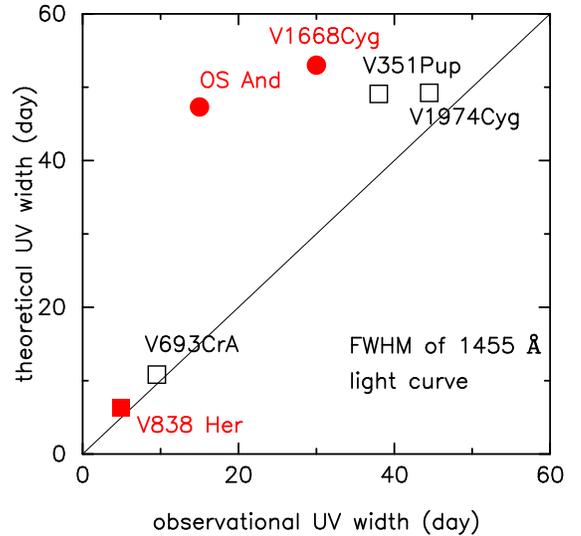}
\caption{Observational and theoretical durations of the UV 1455  \AA~ nova outbursts.
Filled symbols (red) denote novae with dust formation.
Squares and circles denote  Ne and CO novae, respectively. 
Observational data are taken from \citet{cas02}.  The large deviation of OS And and 
V1668 Cyg is due to dust formation. See text for details. 
\label{UVwidthobs.theory}}
\end{figure}

Another possible source of ambiguity in the determination of WD mass may come
from effects of dust.  
Figure \ref{UVwidthobs.theory} compares the duration of UV nova outburst
$t_{\rm FWHM}$, as predicted by our models, with the observed values from
Cassatella et al. (2002) for five novae. The figure shows that there is a
reasonable agreement between the two sets of data, except the dust
forming novae V1668 Cyg and OS And. In these two objects, dust formation took
place shortly after the UV 1455 \AA~ maximum, so making the measured
$t_{\rm FWHM}$ significantly smaller than expected.
In the case of V838 Her, in spite of being a dust forming
nova, the mass of dust was too small to affect the UV light curve, as
discussed in Section \ref{sec:dust}. 

Note that the mass determination of our light curve analysis is based 
not only on the UV data fittings, but also on the optical and IR fittings, so 
the error in UV fitting does not directly reflect to the resultant $M_{\rm WD}$.


Figure \ref{UVwidth} indicates that 
less massive WDs ($\lesssim 1~M_\odot$) are systematically CO-rich, while
more massive WDs ($\gtrsim 1~M_\odot$) are neon-rich. It has been argued
that heavy element enrichment of nova ejecta is caused by dredge-up of WD
core material so that, for example, C and O enrichment is an indication of a
CO WD. An ONe WD is born in massive stars after ignition of a CO core 
and, therefore, ONe WDs are usually more massive than CO WDs when they are 
just born. Our results are consistent with this.

   The mass boundary of CO WDs and ONe WDs are suggested to be 
1.0-1.1 $M_\odot$. \citet{ibe90} summarized that a star of initial mass of 5
-10 $M_\odot$ leaves a CO WD of mass $0.75-1.1 M_\odot$ and a star of $10-12
M_\odot$ becomes ONe WD. \citet{gar97} calculated stellar evolution of 9 
$M_\odot$ star which finally becomes an $\sim 1.08 M_\odot$ ONe core surrounded by
an CO envelope of $ \sim 0.08 M_\odot$. 
\citet{ume99} calculated stellar evolution until CO core ignition with 
fine grid of initial stellar mass and concluded that the mass of CO WDs born 
in binaries is less massive than $1.07~M_\odot$ of which initial mass is 
$\sim 8 M_\odot$ for $Z=0.02$. 
\citet{men08} obtained CO core mass consistent with Umeda et al.'s value.
These predictions are consistent with our results in Figure \ref{UVwidth}.

\subsection{X-ray observations}

X-ray emission from V838 Her has been detected with $ROSAT$ as early as 5
days after the discovery \citep{llo92,obr94}, and was found to be consistent
with emission from shocked plasma with $T \sim 10$ keV \citep{obr94}.
According to these authors the shock could originate from within the ejected
material itself or by interaction of the ejecta with pre-existing
circumstellar matter. In our Model 2 the optically thick wind lasts until
day $\sim 52$~days (see Table \ref{table_model} for other models);  
this is consistent with the picture that the wind collides with
circumstellar matter to produce the observed X-ray flux.


\subsection{Quiescent Phase}

The presence of an accretion disk has been suggested from 
the long duration of the eclipse \citep{ing92}, and from 
the depth of eclipse at minimum \citep{lei93}. 

The absolute magnitude of a disk, seen at an inclination angle $i$, is
approximated by
\begin{eqnarray}
M_{\rm V} {\rm (obs)}=  & &-9.48 -{5\over 3}\log\left({M_{\rm WD}\over M_\odot}
{{\dot M_{\rm acc}}\over {M_\odot ~{\rm yr}^{-1}}} \right) \cr
 & & -{5\over 2} \log(2\cos i),
\label{accretion-disk-Mv}
\end{eqnarray}
where $M_{\rm WD}$ is the WD mass, and $\dot M_{\rm acc}$ is the mass
accretion rate \citep[Eq. (A6) in][]{web87}.  \citet{szk94i} estimated the 
inclination of the disk to be $i=78-90$\arcdeg. Assuming $M_{\rm WD}=
1.35 ~M_{\odot}$ and $i=80$\arcdeg, we get $M_{\rm V} =6.5$, 4.8, and 3.1 for the
mass accretion rates of $1 \times 10^{-9}$, $1 \times 10^{-8}$, and $1 \times
10^{-7} ~M_{\odot}$~yr$^{-1}$, respectively.  The corresponding apparent
magnitudes, as calculated from equation (\ref{modulus}) with $E(B-V)=0.53$
and $d=2.7$ kpc are $m_{\rm V}= 20.3$, 18.6 and 16.9. Considering the ambiguity 
of assumed accretion rates and the inclination angle, 
these values are consistent with the preoutburst
magnitude reported in Section \ref{introduction}.

The contribution of a companion star to the quiescent luminosity is
estimated as follows. \citet{pat84} gives an empirical relation between the absolute
magnitude of a Roche lobe filling companion and its orbital period as
\begin{equation}
M_{\rm V} = 22 - 17.46 \log P ({\rm hr}) ,
\label{companionlum}
\end{equation}
for $0.7 < \log P({\rm hr}) < 1.1$ . For $P ({\rm hr})$ = 7.14 hr \citep{ing92,lei92} 
we get $M_{\rm V}=7.1$ and $m_{\rm V}=20.6$. Thus, the companion star would be fainter than
the accretion disk although there are ambiguity in the mass accretion
rate and the inclination angle.  This is also consistent 
with a faint companion suggested from  a small upper limit of $\sim 0.05$ mag on 
the depth of the secondary minimum \citep{lei93} and from no visible evidence 
in the spectrum \citep{szk94i}.

\section{Conclusions}\label{sec_conclusion}

We have applied the 'universal decline law' of classical novae
described in \S\ref{subsec_universallaw} to one of the fastest novae V838 Her  
and derived various parameters. Our main results are summarized as follows:

\noindent
1. An analysis of the {\it IUE} reprocessed data of V838 Her indicates 
$E(B-V)= 0.53 \pm 0.05$.

\noindent
2. The $1.35 \pm 0.02~M_\odot$  WD model reasonably reproduces 
the light curves of V838 Her both in the optical and in the UV 1455 \AA~ band 
as well as in the infrared $H$ and $K$ bands  at the early stages, 
until dust formation takes place. Model 2 in table \ref{table_model} 
is the best fit model.

\noindent
3. The distance is estimated to be $d \sim 2.7 \pm 0.5$ kpc from the UV 1455
\AA~ light curve fitting.

\noindent
4.  We have estimated ejecta mass $\Delta M_{\rm wind} \sim (2-3) \times 10^{-6}
M_\odot$ lost by winds.

\acknowledgments
The authors are grateful to Takashi Iijima for fruitful discussion, 
and also to an anonymous referee for useful comments to improve the manuscript.
We also thank the American Association of Variable Star Observers
(AAVSO) for the visual data of V838 Her.  This research has been
supported in part by the Grant-in-Aid for Scientific Research
(20540227) of the Japan Society for the Promotion of Science.

\appendix

\noindent

\begin{figure}
\epsscale{.65}
\plotone{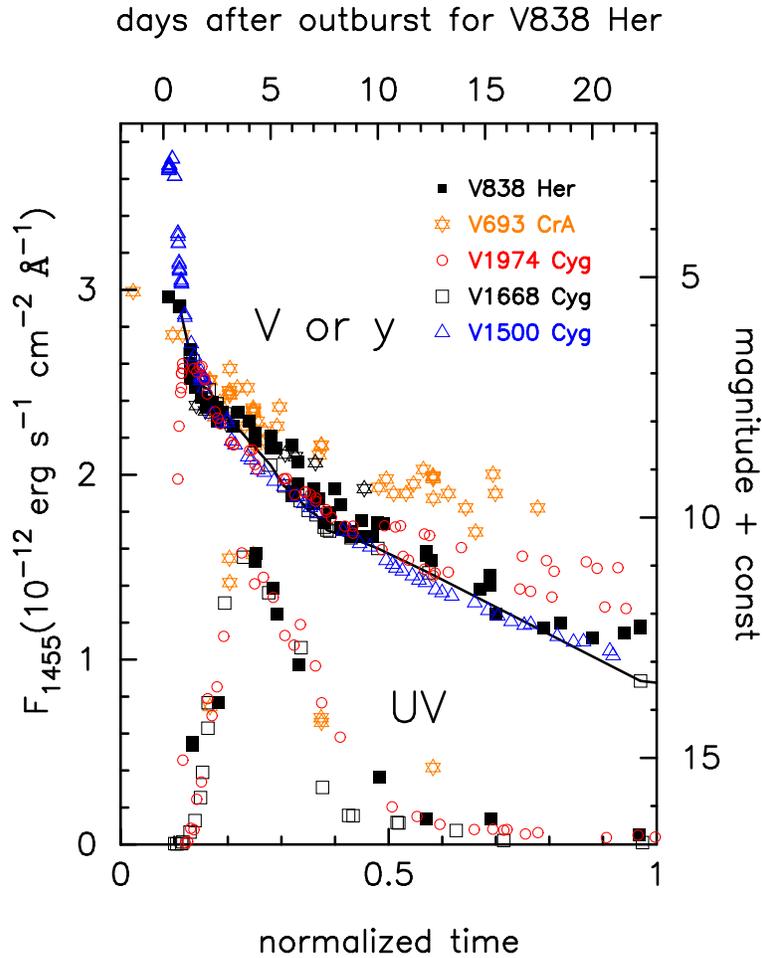}
\caption{ Scaling of five nova light curves. The observed light curve of 
classical novae, V693 CrA, V1974 Cyg, V1668 Cyg, and V1500 Cyg are overlapped to that 
of V838 Her.  For V838 Her only the optical data in IAU Circulars and $IUE$ $V_{\rm FES}$ 
data are shown.
Optical data are $y$ magnitudes for V1668 Cyg (connected by a solid line)
and V1500 Cyg, $V$ for V1974 Cyg (for the references, see Paper I), V and
visual for V693 CrA \citep{kat07h}.  The scale for the UV Flux (on the
left-hand side), the magnitude, and the time after outburst refer
to V838 Her.  The scale of the other novae is normalized as follows: 
the upper limit in the UV flux scale (in
erg~s$^{-1}$cm$^{-2}$\AA$^{-1}$), the lower limit in the magnitude, the
upper limit in the magnitude, and the time normalized to unity
(abscissa) have been set to: (3.9E$-12$, 16.8, 1.8, 25) for V838 Her,
(1.9E$-12$, 18, 3, 54) for V693 CrA, (4.4E$-11$, 14.8, $-0.2$, 202) for V1974 Cyg,
(4.E$-12$, 18, 3, 200) for V1668 Cyg, and (no data, 16.1,1.1, 162.8) for V1500
Cyg.
\label{novacomparison}}
\end{figure}

Figure \ref{novacomparison} demonstrates a ``universality'' of nova light curves,
in which five classical novae are shown in the visual ($V$ and $y$) and UV
1455 \AA~ band (except V1500 Cyg: no UV 1455 \AA~ band is available).  
Each peak of the UV 1455 \AA~ flux is normalized in order to match the peak 
of V838 Her.  Also, the time-scale of each nova is normalized 
so that each UV 1455 \AA~ light curve is overlapped with that of V838 Her. 
The same normalizing factor of time-scale is applied to both the UV and 
the optical light curves. These normalizing factors, obtained by eye 
fitting, are summarized in the figure caption.

Our universal decline law predicts that the light curves, normalized in
this way, should merge into one, independent of the WD mass and chemical
composition (see Paper I).  This match is excellent for the $y$ light
curve, because this band is little contaminated by strong emission
lines \citep{hac08}.  As one may appreciate from the figure, there is a
good overlap between the light curves of V1668 Cyg ($y$-magnitude, connected
by solid line), V1500 Cyg ($y$-magnitude), and the initial 
 phase of V1974 Cyg ($V$ magnitude).  It also
appears from the figure that, whenever strong emission lines provide the
dominant contribution to the magnitude, the data deviate above the V1668
Cyg line. In the case of V838 Her the data are scattered compared with the
above three objects, therefore, we fit its lowest boundary of the data with
the V1668 Cyg line.  We can see excess of the visual magnitudes between day 3
and day 6.

Four UV 1455 \AA~ light curves also show a good agreement with each other,
although V1668 Cyg shows a drop at 0.37 in the normalized time due to dust
formation \citep{geh80}. We don't see any indication of sharp drop like this
in V838 Her.


\begin{thebibliography}{}
\bibitem[Alcock (1991)]{alc91} Alcock, G. 1991, \iaucirc, No. 5222




\bibitem[Cassatella et al. (2002)]{cas02} Cassatella, A., Altamore, A., \&
  Gonz\'alez-Riestra, R. 2002,\aap, 384, 1023

\bibitem[Cassatella et al. (2005)]{cas05} Cassatella, A., Altamore, A., \&
  Gonz\'alez-Riestra, R. 2005,\aap, 439, 205


\bibitem[Cassatella et al. (2004a) ]{cas04a} Cassatella, A., Gonz\'alez-Riestra, 
R., \& Selvelli, P.  2004a, INES Access Guide No.3 Classical Novae (ESA: Netherlands)

\bibitem[Cassatella et al. (2004b)]{cas04b}
Cassatella, A., Lamers, H. J. G. L. M., Rossi, C., Altamore, A.,
Gonz\'alez-Riestra, R. 2004b, \aap, 420, 571

\bibitem[Chandrasekhar et al.(1992)]{cha92}
  Chandrasekhar, T, Ashok, N.M., \& Ragland, S. 1992, \mnras, 255, 412



\bibitem[Della Valle \& Livio (1995)]{del95} Della Valle, M., \& Livio, M.
 \apj, 452, 704

\bibitem[Eggleton (1983)]{egg83} Eggleton, P.P. 1983, \apj, 268, 368

\bibitem[Evans et al. (2003)]{eva03} Evans, A. et al. 2003, \aj, 126, 1981 



\bibitem[Garc\'ia-Berro et al. (1997)]{gar97} Garc\'ia-Berro, E., Ritossa, C.,
\& Iben, I. Jr. 1997, \apj, 485, 765

\bibitem[Gehrz et al. (1980)]{geh80} Gehrz, R.D., Hackwell, J.A., Grasdalen,
  G.L., Ney, E.P., Neugebauer, G., and Sellgren, K. 1980, \apj, 239,570

\bibitem[Grevesse \& Anders (1989)]{gre89}
Grevesse, N., \& Anders, E. 1989, Cosmic Abundances of Matter,
ed. C. J. Waddington (New York: AIP), 1


\bibitem[Groenewegen \& Lamers (1989)]{gro89}
Groenewegen, M. A. T., \& Lamers, H. J. G. L. M. 1989, \aaps, 79, 359





\bibitem[Hachisu \& Kato (2006)]{hac06a}
Hachisu, I., \& Kato, M. 2006, \apjs, 167, 59 (Paper I)

\bibitem[Hachisu \& Kato (2009)]{hac09}
Hachisu, I., \& Kato, M. 2009, \apjl, 694, L103-L106

\bibitem[Hachisu et al. (2006)]{hac06b}
Hachisu, I. et al. 2006, \apj, 651, L141



\bibitem[Hachisu et al. (2008) ]{hac08}
Hachisu, I., Kato, M.,  \& Cassatella, A. 2008, \apj, 687, 1236 (Paper III)




\bibitem[Hachisu et al. (2007)]{hac07kl}
Hachisu, I.,Kato, M., \& Luna, G.J.M. 2007, \apj,  659, L153


\bibitem[Hachisu et al. (1999a)]{hkn99}
Hachisu, I., Kato, M., \& Nomoto, K. 1999a, \apj, 522, 487 





\bibitem[Harrison \& Stringfellow (1994)]{har94} Harrison, T.E. \&
  Stringfellow, G.S.  1994, \apj, 437, 827

\bibitem[Hauschildt et al. (1997)]{hau97}
Hauschildt, P. H., Shore, S. N., Schwarz, G. J., Baron, E., Starrfield, S.,
\& Allard, F. 1997, \apj, 490, 803


\bibitem[Humphreys et al. (1991)]{hum91}
Humphreys, R.M., Zumach, W., \& Stockwell, T. 1991 \iaucirc No. 5224

 
\bibitem[Iben (1990)]{ibe90} Iben, I., Jr. 1990 in Confrontation between stellar 
pulsation and evolution, ed. C. Cacciari, \& G. Clementini, ASP Conf. Proc.
(San Francisco,), 11, 483


\bibitem[Iijima et al. (2009)]{iij09} Iijima,T., Cassatella, A., Kato, M., \& 
   Hachisu, I.  2009, \aap, in preparation

\bibitem[Iglesias \& Rogers (1996)]{igl96}
Iglesias, C. A., \& Rogers, F. J. 1996, \apj, 464, 943


\bibitem[Ingram et al. (1992)]{ing92} Ingram, D., Garnavich, P., Green, P., 
   \& Szkody, P.  1992, \pasp, 104, 402



\bibitem[Kato (1983)]{kat83}
Kato, M. 1983, \pasj,  35, 507


\bibitem[Kato (1997)]{kat97}
Kato, M. 1997, \apjs, 113, 121

\bibitem[Kato (1999)]{kat99}
Kato, M. 1999, \pasj, 51, 525


\bibitem[Kato \& Hachisu (1994)]{kat94h}
Kato, M., \& Hachisu, I., 1994, \apj, 437, 802

\bibitem[Kato \& Hachisu (2005)]{kat05h}
Kato, M., \& Hachisu, I., 2005, \apj, 633, L117

\bibitem[Kato \& Hachisu (2007)]{kat07h} Kato, M., \& Hachisu, I. 2007, \apj,
  657, 1004


\bibitem[Kato \& Hirata (1991)]{tkat91} Kato, T., \& Hirata, R. 1991 \iaucirc  No. 5262

\bibitem[Kato et al.(2008) ]{kat08} Kato, M.,
 Hachisu, I., Kiyota, S., \& Saio, H., 2008, \apj, 684, 1366


\bibitem[Kidger \& Martinez-Roger (1993)]{kid93} Kidger, M.R.  \&
  Martinez-roger, C.  1993, \aap, 267,111



\bibitem[Krautter et al. (1996)]{kra96}
Krautter, J., \"Ogelman, H., Starrfield, S., Wichmann, R.,
\& Pfeffermann, E. 1996, \apj, .456, 788


\bibitem[Lamers et al. (1987)]{lam87}
Lamers, H. J. G. L. M.; Cerruti-Sola, M.; Perinotto, M. 1987,
\apj, 314, 726




\bibitem[Leibowitz et al. (1992)]{lei92} 
Leibowitz, E. M., Mendelson, H., \& Mashal, E.  1992, \apjl, 385, 49 

\bibitem[Leibowitz (1993)]{lei93} Leibowitz, E. M. 1993, \apj, 411, L29


\bibitem[Lloyd et al. (1992)]{llo92} Lloyd, H.M., O'Brien, T.J., Bode, M.F., 
Predehl,P., Schmitt, J.H.M.M., Tr\"umper, J., Watson, M.G., \& Pounds, K.A. 
1992, \nat, 356, 222

\bibitem[Lynch et al. (1992)]{lyn92} Lynch, D. K.,
Hackwell, J.A., \& Russell, R. W. 1992, \apj, 398, 632

\bibitem[Matheson et al. (1993)]{mat93} Matheson,
  T., Filippenko, A.V., \& Ho, L.C. 1993, \apj, 418, L29

\bibitem[Meng et al. (2008)]{men08} Meng, X., Chen, X., \& Han, Z. 2008, \aap,
487, 625

\bibitem[Neckel \& Klare (1980)]{nec80} Neckel, Th., \& Klare, G. 1980, \aaps,
42, 251


\bibitem[Ness et al. (2007)]{nes07}
Ness, J.-U., Schwarz, G. J., Retter, A., Starrfield, S.,
Schmitt, J. H. M. M., Gehrels, N., Burrows, D., \& Osborne, J. P.
2007a, \apj, 663, 505


\bibitem[O'brien et al. (1994)]{obr94} O'brien, T.J.,
  Lloyd, H.M., \& Bode, M.F. 1994, \mnras, 271, 155


\bibitem[Orio et al. (2001)]{ori01}
Orio, M., Covington, J., \"Ogelman, H. 2001, \aap, 373, 542


\bibitem[Patterson (1984)]{pat84} Patterson, J. 1984, \apjs, 54,443

\bibitem[Politano et al. (1995)]{pol95} Politano, M., Starrfield, S., Truran,
  J.W.  Weiss, A., \& Sparks, W.M. 1995, \apj, 448, 807

\bibitem[Prialnik (1986)]{pri86} Prialnik, D. 1986, \apj, 310, 222


\bibitem[Prialnik \& Kovetz (1995)]{pri95}
Prialnik, D., \& Kovetz, A. 1995, \apj, 445, 789


\bibitem[Schmidt (1957)]{sch57}
Schmidt, Th. 1957, Z. Astrophys., 41, 181


\bibitem[Schwarz et al. (2007)]{sch07} Schwarz, G.J., Shore, S. N.,
  Starrfield, S., \& , Vanlandingham, K. M. 2007, \apj, 657,453

\bibitem[Seaton (1979)]{sea79} Seaton, M. J. 1979, \mnras, 187, 73

\bibitem[Smith et al. (1995)]{smi95} Smith, C.H., Aitken, D.K., Roche, P.F., \&
Wright, C.M. 1995, \mnras, 277, 259





\bibitem[Sugano (1991)]{sug91} Sugano, M. 1991, \iaucirc, No. 5222

\bibitem[Stickland et al. (1981)]{sti81}
Stickland, D. J., Penn, C. J., Seaton, M. J., Snijders, M. A. J., \&
Storey, P. J. 1981, \mnras, 197, 107


\bibitem[Starrfield et al. (1992)]{sta92} Starrfield, S., Shore, S. N.,Sparks, 
W.M.,  Sonneborn, G., Truran, J.W., \& Politano, M. 1992, \apj, 391, L71


\bibitem[Szkody \& Hoard (1994)]{szk94h} Szkody, P. \& Hoard, D.W. 1994, \apj, 
429,857 

\bibitem[Szkody \& Ingram (1994)]{szk94i} Szkody, P. \& Ingram, D. 1994, \apj,
  420,830

\bibitem[Ueta (1991)]{uet91} Ueta, E. 1991, \iaucirc, No. 5265

\bibitem[Umeda et al. (1999)]{ume99} Umeda, H., Nomoto, K., Yamaoka, H., \&
  Wanajo, S. 1999, \apj, 513, 861


\bibitem[Vanlandingham et al. (1997)]{van97} Vanlandingham, K. M., Starrfield, S., 
\& Shore, S. N. 1997,  \mnras, 290, 87


\bibitem[Vanlandingham et al. (1996)]{van96} Vanlandingham, K. M., Starrfield,
  S., Wagner, R. M., Shore, S. N., \& Sonneborn, G.  1996, \mnras, 282, 563

\bibitem[West (1991)]{wes91} West, R. M., 1991, \iaucirc No. 5224


\bibitem[Webbink et al.(1987)]{web87}
Webbink, R.F., Livio, M., Truran, J.W., \& Orio, M. 1987, \apj, 314, 653

\bibitem[Williams et al. (1994) Williams, Phillips, \& Hamuy]{wil94}
Williams, R.E., Phillips, M.M., \& Hamuy, M. 1994, \apjs, 90,297

\bibitem[Woodward et al. (1992)]{woo92} Woodward, C.E., Gehrz, R.D, Jones, T.J.,
  \& Lawrence, G.F. 1992, \apj, 384, L41


%
\end{thebibliography}
\end{document}